\documentclass[mathleft]{an}
\usepackage{graphicx}
\usepackage{times}
\usepackage{natbib}
\begin{document}

% The following seven commands are intended for editorial usage and should be ignored by
% the author(s).
\Pagespan{789}{}% Document's page range. 
% If second parameter is left empty, the last page is computed automatically.
\Yearpublication{2006}%
\Yearsubmission{2005}%
\Month{11}%   
\Volume{999}%  
\Issue{88}% 
% \DOI{This.is/not.aDOI}% 

\title{Studying Blazhko RR~Lyrae stars \\ with the 24-inch telescope of the Konkoly Observatory}

\author{\'A. S\'odor\thanks{\email{sodor@konkoly.hu}\newline}}
\titlerunning{Studying Blazhko RR~Lyrae stars}
\authorrunning{\'A. S\'odor}
\institute{Konkoly Observatory of the Hungarian Academy of Sciences. P.O.~Box~67, H-1525 Budapest, Hungary}

\received{30 May 2005}
\accepted{11 Nov 2005}
\publonline{later}


\abstract{About a dozen field RR~Lyrae stars have been observed with the 24-inch Heyde-Zeiss telescope of the Konkoly Observatory at Sv\'abhegy, Budapest, since its refurbishment in 2003. Most of the observing time is allocated for the investigation of the Blazhko modulation, a phenomenon that still does not have a satisfactory explanation. The obtained multicolour CCD observations are unique in extent. The accuracy of the measurements makes it possible to detect low amplitude modulation of the light curve as well. The discovery of Blazhko stars with low modulation amplitudes warns that the incidence rate of the Blazhko modulation is, in fact, much larger than it was previously expected. This makes the efforts exploring the cause of the modulation even more important. A summary of our measurements and results achieved during the last 3 years is presented.}

\maketitle

\section{Introduction}

In this note, our ongoing project of studying Blazh\-ko-mo\-du\-lated fundamental mode RR~Lyrae (RRab) stars with the 24-inch Heyde-Zeiss telescope of the Konkoly Observatory at Sv\'abhegy, Budapest, is introduced. Our first results on the properties of the Blazhko modulation are presented.

\subsection{The Blazhko effect}

The physics of RR~Lyrae stars is considered to be well understood. RR~Lyrae stars play an important role in many fields of astrophysics. Besides being important distance indicators, RR~Lyrae stars are also test objects of stellar evolution and pulsation models. It is a unique advantage of these stars that their fundamental physical parameters can be easily determined, simply from photometric observations of their light curves \cite[]{vas,fund,kovacs}. Our knowledge on these variables is summarized by \cite{smith}.

There are ``normal", monoperiodic RRab stars with re\-mar\-kably stable light curves (see Jurcsik et al. 2006b for examples) and there are modulated ones. Two types of multi-periodic RR~Lyrae stars are known: the Blazhko-mo\-du\-la\-ted and the double-mode RR~Lyrae variables (about double-mode pulsation, see D\'ek\'any 2007, in this issue). The pheno\-menon of variation in the phase and brightness of the maximum light was discovered by Blazhko (1907, phase modulation of RW~Dra) and Shapley (1916, amplitude modulation of RR~Lyr) a century ago, and it is called Blazhko effect today. This variation shows periodic behaviour on days to hundred days time scale.

\begin{figure}
  \begin{centering}
    \includegraphics[width=82mm]{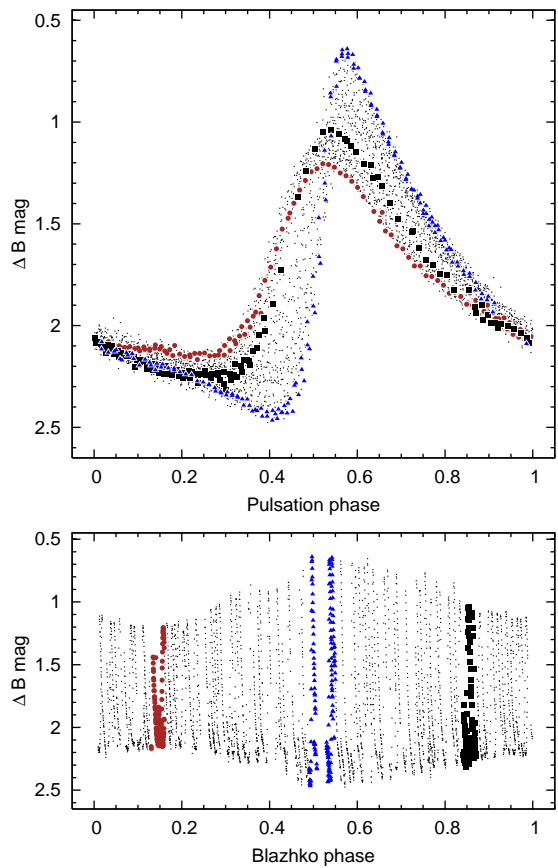}
    \caption{The modulated light curve of MW~Lyrae. The upper panel shows the $B$ light curve phased with the pulsation period, while the same data are shown versus the Blazhko phase in the lower panel. Three different Blazhko phases around minimum, maximum, and mean amplitude phases are highlighted.}
    \label{fig:blmod}
  \end{centering}
\end{figure}

As an example for the light curve modulation, the panels in Fig.~\ref{fig:blmod} show the measurements of MW~Lyrae\footnote{The observations of MW~Lyrae were made in collaboration with H. A. Smith, Michigan State University, with the 24-inch Heyde-Zeiss telescope, Budapest and with the 60-cm telescope of the MSU Campus Observatory, East Lansing, in 2006.} phased with the pulsation and modulation periods, respectively. Three different Blazhko phases are highlighted. This variable has a particularly strong and stable modulation.

\subsubsection{Photometric investigations}

Utilizing photometric observations the Blazhko effect can be studied through $O-C$ data and maximum brightness values, or, if the phase coverage of the observations permits, by Fourier analysis of the complete light curve. The $O-C$ data of the timings of the maximum light or of a specific point of the light curve show the periodicity and the strength of the phase modulation. The amplitude modulation can be studied through the maximum brightness -- maximum timing data similarly. Plotting the maximum brightness versus the maximum timing $O-C$ data, we obtain a diagram showing a loop which characterizes the modulation of the variable (see e.g. Fig. 5 in Jurcsik et al. 2006a). The properties of the Blazhko modulation of individual stars often change with time, either in a quasi-regular or in an irregular way.

In 2004, at the beginning of our project, there were only about a dozen Blazhko stars with extensive photometric observations sufficient for studying the Blazhko modulation in details. The occurrence rate was estimated to be about 20-30\% among RRab stars (Szeidl 1988, Alcock et al. 2003, Moskalik \& Poretti 2003).
\vskip 5mm

The Blazhko modulation can also be investigated spectroscopically which is a fruitful area of research. The brightest Blazhko star, RR~Lyrae itself, is thoroughly studied in this way by M. Chadid and her coworkers (Chadid \& Chapellier 2005, and references therein).

\subsubsection{Theories}

There are 3 different ideas about the physical origin of the Blazhko effect, however, none of them explain satisfactorily all the observed aspects of the century-long known phenomenon. Two theories involve excitation of nonradial pulsation modes and relate the modulation period to the rotation of the star. Indirect evidence of the connection between the  modulation and rotation periods of Blazhko stars has been found by \cite{acta}.

One of these theories supposes nonlinear resonant coupling of the radial mode and a low order nonradial one. Typically the azimuthal order of the nonradial mode is 1 and the rotational splitting causes the modulation \cite[]{nowak}. The problems with this model are summarized by \cite{dzm}.

The oblique magnetic rotator model is another possible explanation of the Blazhko modulation. This model is applied successfully for roAp stars. Here a bipolar magnetic field is assumed whose axis inclines to the axis of the rotation. This field has to be strong enough to distort the radial pulsation in such a way that a nonradial mode with horizontal degree of 2 is excited. As the star rotates our aspect angle changes and we observe different light curve shapes \cite[]{cousens,shiba}. One argument against this Blazhko theory is that it predicts quintuplet structure of the Fourier spectrum of the light curve, but this was never observed.

The third theory assumes periodic changes in the structure of the convective envelope of the star. The turbulent convection in the envelope varies due to a changing magnetic field generated by turbulent or rotational dynamo \linebreak mechanism. The gradual strengthening and weakening of the field happens periodically which leads to light curve modulation. This idea has been put forward very recently by \cite{stothers}. It is known that turbulent dynamo causes cyclic variations in the magnetic field, however, the Blazhko modulation usually shows more regular, periodic behaviour, which would be difficult to explain with stochastic physical processes.

\section{About our project}

\subsection{Telescope}

The 24-inch Heyde-Zeiss telescope of the Konkoly Observatory at Sv\'abhegy, Budapest, was refurbished and automated in 2003. Thanks to the automatisation, the telescope can be operated by remote control, with low human re\-sour\-ces. The telescope is operated by undergraduate and PhD students in astronomy. The weather conditions of this site make it possible to do observations on about the half of the nights.

\begin{figure}
  \begin{centering}
    \includegraphics[width=70mm, height=70mm]{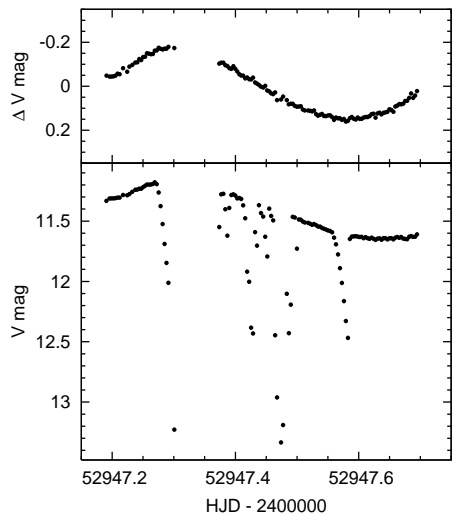}
      \caption{The power of the differential photometry. Variation in the sky transparency causes dimmings in the observed instrumental brightness (bottom panel), which do not affect the differential light curve significantly (top panel). The plots show observations of V823~Cas \cite[]{v823}.}
    \label{fig:difflc}
  \end{centering}
\end{figure}

Despite the bright sky of Budapest, differential photometric measurements with standard Johnson and Cousins $B$, $V$, $R_\mathrm C$, and $I_\mathrm C$ filters are obtained with a typical accuracy of 0.01\,mag for stars between 10th and 14th magnitudes. Fig.~\ref{fig:difflc} demonstrates the insensitivity of the differential photometry to the variation of the sky transparency. Even a more than 1.5\,mag dimming in the observed instrumental brightnesses does not affect the differential magnitudes significantly.

About 80\% of the observing time is dedicated to the investigation of the Blazhko effect. 

The advantage of this telescope is that without time li\-mit, we can obtain extended multicolour observations of many Blazhko stars with good photometric accuracy.

About the telescope and the programs see further details at http://www.konkoly.hu/24/.

\subsection{Our survey}

As there were no satisfactory explanation of the modulation and because very few Blazhko stars had extensive enough photometry, we initiated a survey of brighter ($V < 14\,\mathrm{mag}$ at minimum light), short period ($P < 0.5\,\mathrm{d}$), fundamental mode RR~Lyrae stars of the northern sky in 2004. This survey makes it possible to refine the incidence rate of the modulation and provides uniquely extended multicolour light curves of Blazhko stars to study the details of the Blazhko effect. Fig.~\ref{fig:blmod} demonstrates the extensiveness of our observations. It shows the $B$ light curve of MW~Lyrae which variable was observed in $B$, $V$, and $I_\mathrm C$ bands. The light curves contain more than 3600 measurements of 116 nights in each band. The data can be divided into 20 parts according to the Blazhko phase so that the pulsation cycle is covered in each part completely.

\section{First results}

\subsection{Weakly modulated stars}

The first target of this project was RR~Geminorum \citep{rrgem1} and it had beaten two records immediately. RR~Gem proved to be modulated but with the lowest amplitude ($A_{V\,\mathrm{max}} = 0.09\,\mathrm{mag}$) in maximum brightness variation and with the shortest period ($P_\mathrm{mod}=7.2\,\mathrm d$) known at that time. Later we have found SS~Cancri also to be modulated \citep{sscnc} with the even shorter period of 5.3\,d and with similarly low modulation strength.

\subsection{Multi-periodic modulation}

\begin{figure}
  \begin{centering}
    \includegraphics[width=83mm,height=70mm]{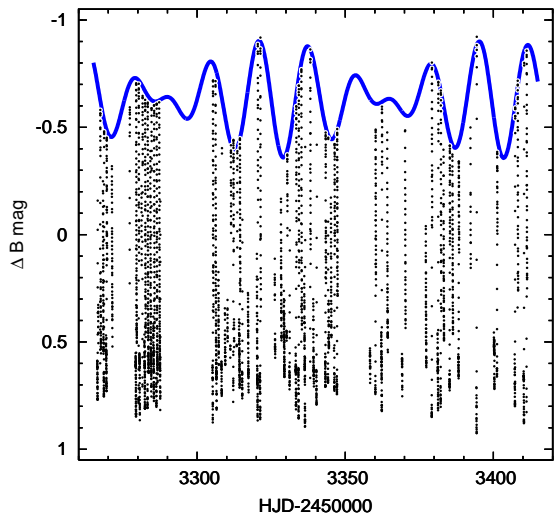}
      \caption{Light curve of CZ~Lac from the 2004-2005 observing season. The upper curve shows the envelope of the maximum brightnesses. The beat effect in this curve is caused by the closeness of the modulation periods and their similar strength.}
    \label{fig:czlac}
  \end{centering}
\end{figure}

We have found RR~Lyrae stars showing multi-periodic modulation. The light curve variation of these stars cannot be satisfactorily described with only one modulation period. In the case of UZ~Ursae~Majoris, two very different modulation periods were needed to fit the data \citep{uzu}. These peri\-ods differ with a factor of more than 5 (26.7\,d and 143\,d). The multi-periodic modulation was suspected earlier by \cite{lacluyze} in the case of XZ~Cyg. 
Non-equidistant triplets have been found in the Fourier spectra of several RRab stars of the MACHO \cite[]{macho} and OGLE \cite[]{ogle} databases, which can also be a sign of multi-periodic modulation. These observations, however, do not allow a detailed study of the complex modulation of these stars. UZ~UMa provided the first unambiguous evidence of the existence of the multi-periodic Blazhko modulation.

CZ~Lacertae is another fundamental mode RR~Lyrae variable which we have found to be modulated and also with two periods. Both modulations have similar strength so there is no dominant modulation period. Fig.~\ref{fig:czlac} shows our observations of this star from the 2004-2005 observing season with a two component harmonic curve fitted to the maximum brightnesses. It displays a strong beat effect due to the closeness of the two modulation frequencies. The periodic decrease of the modulation amplitude resembles the cessation of the modulation of RR~Lyrae in about every 4th year (Fig.~6 in Szeidl 1976). In the case of CZ~Lac, as a result of the interaction of the two close modulation frequency components, the strength of the modulation (with about a 18\,d long period) varies with a longer beat period ($\approx$75\,d). Maybe the 4-year long cycle of RR~Lyrae is also a manifestation of the interaction of two closely spaced modulation periods of about 39-41 days. To resolve these peaks in the Fourier spectrum, long (more than about 6 years) and continuous observations are needed. Though RR~Lyrae is the most extensively studied Blazhko variable, its frequent and irregular pulsation period changes (Szeidl \& Koll\'ath 2000) render the Fourier spectrum of such a long light curve hard, if not impossible, to interpret, and hinder the detection of the multiple periodicity of the modulation. The simultaneous existence of different modulation periods of RR~Lyrae may explain the various reported Blazhko period values (between 38.8\,d and 42\,d) published by different authors (Table~6 in Kolenberg et al. 2006, and references therein).

The detection of the multi-periodic modulation warns that the modulation period is not a unique property of these stars.

\section{Conclusion}

In the course of this project, we have observed 12 fundamental mode RR~Lyrae stars so far. Three of them were supposed to have light curves stable enough to be used for the calibration of the formulae which express the physical parameters of the variables from the parameters of the light curves \citep[TZ~Aur, SS~Cnc, and RR~Gem;][]{vas}. Eight targets had no detailed photometry previously (BH~Aur, BK~Cas, EZ~Cep, SS~Cnc, CZ~Lac, TW~Lyn, ET~Per, BR~Tau, UZ~UMa) and two variables were suspected to be modulated, but the observations were contradictory (RR~Gem: Bal\'azs-Detre 1960, Det\-re 1970, Jurcsik et al. 2005a; MW~Lyr: Gessner 1966, Mandel 1970, S\'odor \& Jurcsik 2005).

Surprisingly, we have found that half of these 12 stars are unambiguously modulated. Three Blazhko stars show only small amplitude modulation, where the variation in maximum brightness is less than 0.1\,mag in $V$ band. The other 3 have large amplitude modulation. Furthermore, 3 of the 6 Blazhko stars show multi-periodic modulation.

The existence of the multi-periodic modulation poses a great challenge against any theoretical model of the Blazhko effect, especially against those which relate the modulation period to the rotation of the star.

The modulation seems to be a more common feature of RR~Lyrae stars than it was suspected earlier. We have found RR Lyrae variables with such a weak modulation that was not known earlier. To detect this kind of modulation, extended and accurate enough CCD or photoelectric observations are needed. The mass photometry projects like MACHO, OGLE, ASAS, NSVS and others do not provide light curves appropriate for this detection. Accordingly, we can state that previous statistics were most probably based only on large modulation amplitude variables. It seems that the frequency of the modulated stars increases as the accuracy and the extension of the observations is enhanced. Therefore, based on the available data, it is hard to estimate the real incidence rate of the modulation. Anyway, it seems plausible that at least half of the fundamental mode RR Lyrae variables are modulated, and it cannot even be excluded, that the modulation is a common property of every variable of this type and that modulation with only a hundredth of a magnitude also exists.

Consequently, we cannot state that we know RR~Lyrae type variables really well until the puzzle of the Blazhko effect is solved. This makes the efforts exploring the cause of the modulation even more important.

\acknowledgements I would like to express my gratitude to every member of this project, in particular to J. Jurcsik, B. Szeidl and M. V\'aradi. These results could not be achieved without their joint effort. I thank for J. Jurcsik and B. Szeidl also for their useful suggestions during the elaboration of this paper. I thank the referee, \linebreak L. Szabados for his helpful comments. I would also like to thank H. A. Smith, Michigan State University, for the collaboration in the observations of MW~Lyrae. The financial support of OTKA grants T-043504, and T-048961 is acknowledged.

\end{document}